# Design and preliminary operation of a laser absorption diagnostic for the SPIDER RF source

M. Barbisan[a,b], R. Pasqualotto[b] and A. Rizzolo[b]

[a] *INFN-LNL, v.le dell'Università n. 2, I-35020, Legnaro (PD) Italy*

[b] *Consorzio RFX, C.so Stati Uniti 4, I -35127, Padova, Italy*

The ITER Heating Neutral Beam (HNB) injector is required to deliver 16.7 MW power into the plasma from a neutralised beam of H⁻/D⁻ negative ions, produced by an RF source and accelerated up to 1 MeV. To enhance the H⁻/D⁻ production, the surface of the acceleration system grid facing the source (the plasma grid) will be coated with Cs because of its low work function. Cs will be routinely evaporated in the source by means of specific ovens. Monitoring the evaporation rate and the distribution of Cs inside the source is fundamental to get the desired performances on the ITER HNB. In order to proper design the source of the ITER HNB and to identify the best operation practices for it, the prototype RF negative ion source SPIDER has been developed and built in the Neutral Beam Test Facility at Consorzio RFX. A Laser Absorption Spectroscopy diagnostic will be installed in SPIDER for a quantitative estimation of Cs density. By using a wavelength tunable laser, the diagnostic will measure the absorption spectrum of the 852 nm line along 4 lines of sight, parallel to the plasma grid surface and close to it. From the absorption spectra the line-integrated density of Cs at ground state will be measured. The design of this diagnostic for SPIDER is presented, with details of the layout and of the key components. A preliminary installation of the diagnostic on the test stand for Cs ovens is also described, together with its first experimental results; the effect of ground state depopulation on collected measurements is discussed and partially corrected.

## 1. Introduction

The additional heating systems of the ITER reactor include Neutral Beam Injectors (NBI) capable of producing a beam of hydrogen/deuterium particles accelerated at 870keV/1MeV, respectively, for an overall power of 16.7 MW/beam [1]. At present, the most efficient way to produce $H^0/D^0$ particles at the required energy is to neutralize a beam of $H^-/D^-$ ions by means of stripping reactions with the gas molecules inside the HNB. To produce negative ions, it was chosen to use an Inductively CouPled (ICP) RF plasma source, coupled to the first grid of the extraction and acceleration system, the Plasma Grid (PG). Most of negative ions are produced by means of a surface reaction on the PG, which is covered with Cs, a low work function metal. Cs will be deposited on the PG by evaporation from multiple Cs ovens installed in the source. From the operation of similar sources [2-4] it has emerged that the distribution of caesium inside the source, also due to the plasma action on the surfaces, has a deep impact on the production efficiency, time stability and uniformity of the beam. The thickness of the Cs layer on the PG surface can alter the work function and then the ion production rate.

The development of the ITER HNBs foresees the construction of a prototype of the source with a 100 keV acceleration system: SPIDER) [5]. The source is a part of the neutral beam test facility at Consorzio RFX (Padua) and entered into operation in June 2018. Optimizing the Cs evaporation and the plasma erosion in order to have an optimal and stable layer of Cs will clearly be one of the most important and difficult targets. In order to monitor the amount of Cs in several points in the source, the emission spectrum of Cs will be studied by means of the Optical Emission Spectroscopy (OES) diagnostic [6]. Whereas the Cs density cannot be straightforwardly estimated from the emission spectral lines (because influenced by the plasma conditions, e.g. electron temperature and density), the Cs density can be more directly measured with a laser absorption spectroscopy (LAS) diagnostic [7].

Sec. 2 will briefly summarize the principle of operation of the LAS diagnostic, while Sec. 3 will present the design of this diagnostic for the SPIDER source. Sec. 4 will then present the preliminary results obtained by a similar diagnostic, installed in a test stand dedicated to the experimentation on the Cs ovens for SPIDER (CATS – CAesium oven Test Stand) [8]. The effect of ground state depopulation on the measurements is characterized and partially corrected in Sec. 5.

## 2. Principle of operation of the diagnostic

Essentially, a LAS diagnostic consists of a finely tunable laser, whose light beam crosses the source parallel to the PG and is then collected on the opposite side by a collimator. The intensity of the received light is then

_________________________________________________

*author's email: marco.barbisan@igi.cnr.it*

measured by a photodiode. The signal of the photodiode is finally amplified and digitized. During the diagnostic operation, the laser emission wavelength is linearly varied, repetitively scanning a wavelength interval comprising the Cs $D_2$ 852.1 nm line ($6^2P_{3/2}$–$6^2S_{1/2}$). From the scans the absorption spectrum of the Cs atoms at ground state is obtained. Being the laser spectral width of few MHz, the $D_2$ line is actually resolved in two lines, each one including three transitions: F=3→F=2,3,4 and F=4→F=3,4,5. The two lines are separated by 21.4 pm [7]. From each absorption spectrum, the line-integrated measurement of Cs density at ground state $n_k$ can be obtained [7]:

$$n_k = \frac{8\pi c}{\lambda_0^4} \frac{g_k}{g_i} \frac{1}{A_{ik}l} \int ln\left[\frac{I(\lambda,0)}{I(\lambda,l)}\right] d\lambda \qquad (1)$$

where $c$ is the speed of light, $\lambda_0 = 852.11$ nm is the $D_2$ line wavelength, $g_k = 2$ and $g_i = 4$ are the statistical weights of lower and upper level, $A_{ik} = 3.276 \cdot 10^7\ s^{-1}$ is the transition probability for spontaneous emission, $l = 0.87\ m$ is the length of the laser beam path inside the SPIDER cesiated source volume, $I(\lambda,0)$ and $I(\lambda,l)$ are the intensity of the laser beam at wavelength $\lambda$ before and after having passed through the cesiated volume. The measurement can be critical if the laser light is too low or too high: in the first case, the absorption lines are saturated; in the second case, the absorption of the laser photons leads to the depopulation of Cs at ground state, then to a systematic underestimation of $n_k$.

## 3. Design of the diagnostic for SPIDER

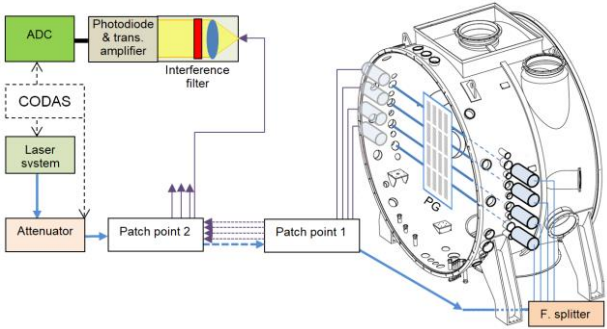

Fig. 1. Scheme of the LAS diagnostic in SPIDER.

For SPIDER, the LAS diagnostic has been designed as briefly sketched in fig. 1. The laser light will be produced by a 852 nm Distributed FeedBack (DFB) diode (*Sacher Lasertechnik* DFB-0852-150-TO3), whose emission wavelength can be linearly tuned by varying its temperature (0.06 nm/°K) and the current flowing in it (3pm/mA). The diode will be managed by a dedicated controller (*S.L.* Pilot PC 500), connected to the SPIDER Control and Data Acquisition System (CODAS). In particular, a board (*National Instruments* NI PXIe-6259) will generate sawtooth voltage signals (16 bit resolution, min. 1.25 MS/s) for the laser controller to repetitively and linearly modulate the laser current and then emitted wavelength and laser intensity.

The laser head hosting the DFB diode will provide for protection from back reflections (35 dB isolation) and for coupling the laser output into the FC/APC connector of a 5μm(core)/125μm(cladding) Single Mode fiber, with 0.11 numerical aperture. Ideally, with no current modulation 60 mW laser output power should be available in fiber. The downstream device will be a max. 30 dB fiber variable attenuator (*Thorlabs* V800A), voltage controlled by CODAS, to vary the laser output power and study the effect of depopulation on the Cs density measurements. The output of the attenuator will be connected to a 7 m long single mode fiber (5μm/125μm, with FC/APC connectors), up to the patch point 2. All these devices will be installed in the SPIDER diagnostic control room. From patch point 2 another SM fiber, 21 m long, carries the light to patch point 1, located inside SPIDER's bioshield. Another SM fiber, 11 m long, will carry the laser light on the opposite side of SPIDER. There the laser output will be equally split among 4 fibers, 5 m long, to reach the collimators on the viewports of SPIDER vacuum vessel.

The collimators will emit 3.6 mm wide beams, which will enter the vessel and pass through the 10 mm wide lateral apertures in the source walls. All the laser beams will pass parallelly to the PG. To extract and accelerate the negative ions, this grid and the other ones have 1280 apertures, positioned in 16 groups which are in turn arranged in 4 columns and 4 rows (fig.1). Each laser beam will pass in front of a row of aperture groups to study the vertical asymmetry in Cs density, which is expected to be the most relevant one, given the direction of the magnetic filter field in the SPIDER source [5]. The distance of each laser beam from the PG can be varied from 5 mm to 65 mm in order to further spatially characterize the distribution of Cs. Lines of sight are available also to check the presence of caesium in the gaps between the 3 grids of the acceleration system. Past a 87 cm long path in the plasma, the laser beam will exit the source from specular apertures and then reach the viewports on the opposite side of the vacuum vessel.

The 4 laser beams will be collected by as many collimators, with an optical aperture of 10 mm and mounting a plano-convex lens of 15 mm diameter and 120 mm focal length. The light will be focused on the FC/PC connectors of 1000μm/1100μm Multi-Mode fibers. Patch chords of 11 m, 21 m and 7 m will connect the collimators to the two patch points and finally to the detectors, back in the diagnostic control room. The light exiting the fibers will be focused by an aspheric lens with 10 mm diameter and 8 mm focal length to the 16 mm² active area of a photodiode [9]. Between lens and photodiode, an interference filter (*Andover C.* 064FSX10-12.5) with 10 nm FWHM bandwidth at 852 nm will stop spurious light collected from the plasma. The photodiode (*OSI* OSI020-UV) will have a responsivity of about 0.55 A/W at the desired wavelengths, and will embed an operational

_________________________________________________

*author's email: marco.barbisan@igi.cnr.it*

amplifier with 5 MHz gain-bandwidth product, used as transimpedance amplifier ($1\cdot10^5$ V/A or $5\cdot10^6$ V/A). The output voltage signal will then be optionally filtered and amplified with a gain 1/10/100/1000. For optimal synchronization, the 4 output signals will be acquired by the same CODAS board which produced the modulation signal for the laser. The signals will be acquired at max. 0.25 MS/s multiplexed or 1.25 MS/s if single channel, with 16 bit resolution, within 7 possible voltage ranges from ±0.1 V to ±10 V.

## 4. Preliminary measurements

Part of the instrumentation already procured for the LAS diagnostic in SPIDER was used to assemble a similar diagnostic in CATS, the test stand for the Cs ovens to be installed in SPIDER. The target was to monitor the evaporation of Cs from the oven installed in the test stand and contemporarily validate the design of the diagnostic and its components. Figure 2 shows the vacuum vessel hosting the Cs oven. The 852 nm laser beam is injected by the input collimator at the top and collected by the reception one at the bottom. The Kodial® windows dedicated to the collimators have been protected from the direct Cs flow from the oven nozzle, by installing them at the end of 12.5 cm long vacuum pipes. The laser path in vacuum is 80 cm long, comparable to the laser path in the plasma of the SPIDER source. Compared to the design of the LAS diagnostic for SPIDER, the only differences in this case are the reduced length of the fibers (7 m), the increased diameter of the laser beam (6.5 mm) due to the use of a different input collimator and the positioning of the interference filter which was embedded in the reception collimator. The detection device is also different: *Thorlabs* PDA-36A-EC, embedding a Si photodiode with 0.55 A/W responsivity at 852 nm and a variable gain transimpedance amplifier, from $7.5\cdot10^2$ V/A to $4.8\cdot10^6$ V/A (bandwidth from 5 MHz to 5 kHz).

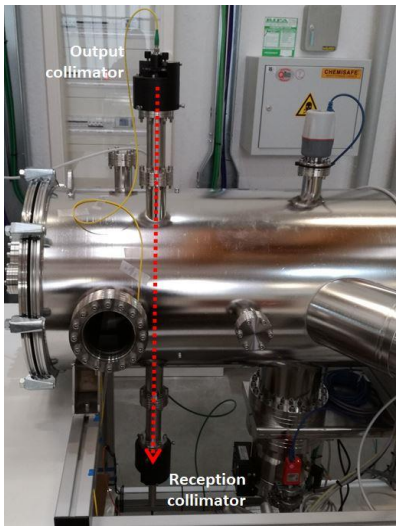

Fig. 2. View of the LAS transmission (top) and reception (bottom) collimators in the CATS test bed.

Figure 2 shows the photodiode's signal during a linear scan in laser intensity and wavelength, obtained by

author's email: marco.barbisan@igi.cnr.it

the LAS diagnostic in CATS during Cs evaporation. Ideally, the signal should consist of a ramp, interrupted by the 2 absorption peaks composing the Cs $D_2$ line. Actually, the signal is distorted by slow quasi-periodic oscillations. Dedicated tests showed that the cause is a phenomenon occurring in one or more devices in between the two collimators, probably etaloning. The presence of the interference filter resulted to partially contribute to this phenomenon, therefore the filter was removed, as this just tolerably increases the background signal.

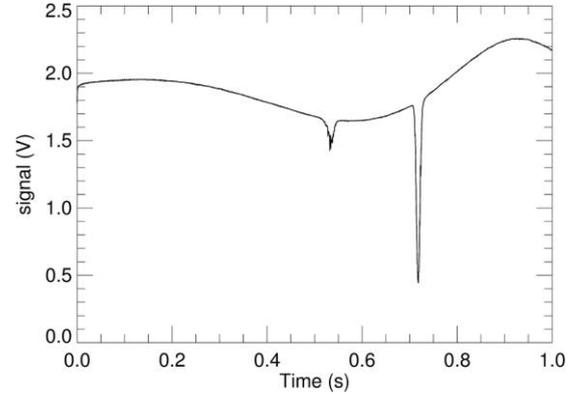

Fig. 3. Typical signal of an absorption spectrum of the Cs $D_2$ line, as acquired by the LAS diagnostic in CATS.

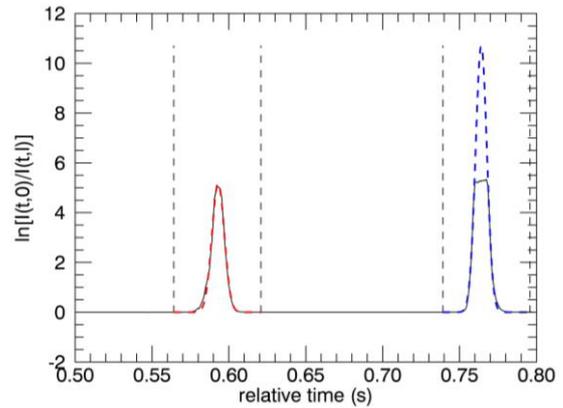

Fig. 4. Saturated spectrum of the Cs $D_2$ line, in terms of logarithm of the ratio between the laser intensity before and after the absorption caused by Cs. The red and blue dashed lines indicate the Gaussian fits on the absorption peaks. The fit intervals are delimited by vertical dashed lines.

In order to estimate the Cs density as in Eq. (1), it is necessary to integrate the logarithm of the ratio between the laser intensity before and after the travel through the vacuum region. The former signal cannot be measured, but can be estimated from the latter, which is represented by the photodiode's signal. A 9$^{th}$ degree polynomial fit is applied to a single wavelength scan; from its residuals the positions of the two absorption peaks are identified. ±3% of the signal length is removed around each peak, and the polynomial fit is repeated. The polynomial curve is then used as an estimation of the laser intensity upstream the vacuum volume. Figure 4 shows an example of the logarithm of the ratio between the two mentioned signals, in a case in which one absorption peak is saturated. The flattop is caused by broad spectral intrinsic baseline of the laser emission. This emission is collected at any point of the scan, resulting in an offset of the photodiode's signal which in turn causes a clipping of $ln[I(t,0)/I(t,l)]$ [7,10]. To cope this issue, the formerly excluded signal intervals including the absorption peaks (indicated in Fig. 4 with dashed vertical lines) are fitted with Gaussian curves (plotted as red and blue dashed lines). The points in the saturation flattop are excluded by imposing an inferior limit $V_{th}$ to the voltage of the original signal (5 mV in the case of fig. 3). After the fit, the Gaussian curves are rescaled from a temporal base to a wavelength base, given the known wavelength interval between the absorption peaks. The integral of eq. 1 and then the Cs density can be finally estimated. In the case of Fig. 4, the estimated density is $5.2 \cdot 10^{15}$ m$^{-3}$. The adoption of the $V_{th}$ threshold is useful also in those cases in which the signal derivative is too high for the detector bandwidth; in such situation, the voltage signals could undergo an undershoot effect at the absorption peaks, with a consequent deformation of the apex of the peaks. The issue can be solved by choosing an adequate value of $V_{th}$ and lowering the wavelength scan interval and the detector's amplification.

During the experimental campaign in CATS, it resulted that the minimum detectable density is around $10^{13}$ m$^{-3}$, while the maximum measurable density is in the order of $10^{16}$ m$^{-3}$, depending on line saturation conditions. Regarding the measurement uncertainties, the standard deviation of Cs density measurements is around 2%-5%, but can reach 10%-15 % in case of line saturation. The main source of error is however likely the ground state depopulation systematic effect. Tests and simulations are ongoing to quantify this effect and properly correct the measurements.

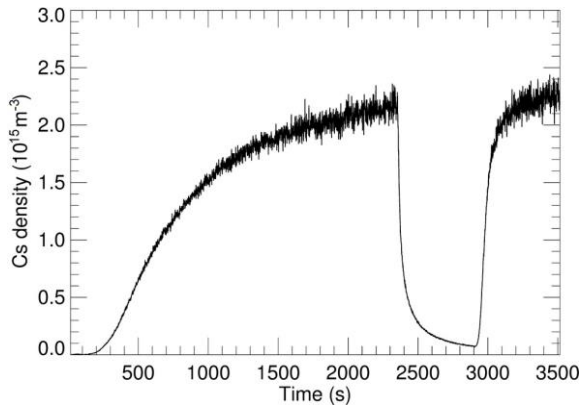

Fig. 5. Temporal evolution of the Cs density in CATS, as measured by the LAS diagnostic during two subsequent openings of the Cs oven valve.

An example of Cs density measurements is given in fig. 5, showing the temporal evolution of the Cs density in CATS during the opening, closure and subsequent re-opening of the main valve of the Cs oven [8]. Before and during the measurements the Cs reservoir of the oven was kept at 150 °C, while the duct connecting the reservoir to the exit nozzle was kept at 200°C.

________________________________________

*author's email: marco.barbisan@igi.cnr.it*

## 5. The depopulation effect

A substantial source of systematic error which may affect the Cs density measurements is given by the depopulation of the ground level: at high laser beam intensity or at low Cs density, it must be taken into account that a not negligible fraction of the Cs atoms is in an excited state. The consequence is an underestimation of the Cs density. Figure 5 shows 13 measurements of the LAS diagnostic in CATS, performed at constant Cs density but different laser beam intensities, obtained by acting on the fiber variable attenuator. The uncertainties on the measurements are about 2 % in Cs density and 15 % on laser beam intensity. As shown by Fig. 5, the effect of depopulation significantly affects the measurements with laser intensities down to about 1 W/m$^2$. The analysis of the absorption spectra showed that the depopulation effect noticeably modifies the ratio between the integrals of the peaks in the spectra plotted in Fig. 3-4. Figure 5 shows the ratio of the integrals of the 2 peaks, the one at lower wavelength (F=3→F=2,3,4) over the one at higher wavelength (F=4→F=3,4,5), again as function of the laser beam intensity; the uncertainty on the values of the ratio is about 5 %. It is evident that with higher laser intensity the peak at lower wavelengths (on the left in Fig. 3-4) becomes weaker compared to the other one. Evidently, the balance between the reactions of excitation and subsequent spontaneous/stimulated emission tends to disfavor the population of the ground level $6^2S_{1/2}$ F=3. This evidence can be used as a further means to detect ground state depopulation.

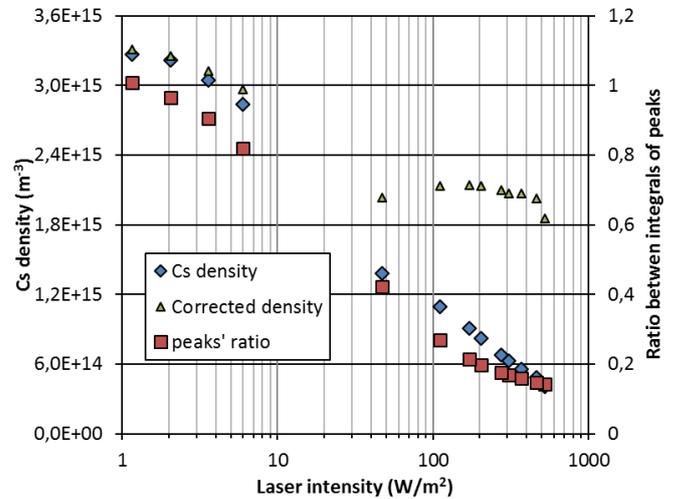

Fig. 5. Cs density measurements (blue diamonds) as function of laser beam intensity; the density values corrected according to the depopulation model are indicated with green triangles. The plot shows also the ratio of the integrals of the two peaks (low wavelength over higher wavelength), with red squares, again as function of laser beam intensity.

Efforts have been made in order to correct the effect of depopulation from the present data; the issue may be directly avoided by keeping the laser intensity low enough, however this may not be always feasible, if the Cs density were too high and the line saturation were impacting too much on the line shape. Considering that the natural linewidths for the single transitions involved in the Cs D$_2$ line are below 0.07 pm, and that the Doppler broadening at room temperature for Cs at 852 nm is one order of magnitude larger (0.9 pm FWHM), the ground state depopulation could be treated according to the model of the inhomogeneous Doppler broadening [11]. Assuming that the velocity distribution of Cs atoms follows a Maxwellian distribution, the absorption coefficient $k_0 = -(dI/dl)/I$ is attenuated to

$$k = \frac{k_0}{\sqrt{1+\beta I}}, \quad \beta = \frac{2\sigma}{h\nu A_{ik}} \quad (2)$$

where $\sigma = 2.3 \cdot 10^{-13}$ m$^2$ is the absorption resonant cross section [12], $h$ is the Planck constant and $\nu$ is the electromagnetic wave frequency. With this correction, Eq. 1 can be rewritten as follows [13]:

$$n_k = \frac{8\pi c}{\lambda_0^4 A_{ik} l} \frac{g_k}{g_i} \int \{f[I(\lambda, 0)] - f[I(\lambda, l)]\} d\lambda \quad (3)$$

$$f(I) = 2\sqrt{1+\beta I} + \ln\left(\frac{\sqrt{1+\beta I}-1}{\sqrt{1+\beta I}+1}\right) \quad (4)$$

To perform the calculation, the photodiode output signal was calibrated to be converted in W/m$^2$ units. As example, Fig. 5 shows the corrected measurements of Cs density (green triangles) obtained by the absorption spectra collected at different laser beam intensity. As it can be seen, the difference between the measurements at high laser intensity and the supposed undeviated value (3.2·10$^{15}$ m$^{-3}$) is clearly reduced but still unacceptable. A more accurate correction would probably need to take separately into consideration the hyperfine structure of the Cs D$_2$ line, since it is evident from the two peak integrals ratios in Fig. 5 that the $F = 3$ and $F = 4$ ground hyperfine levels are not depopulated in the same way.

## 6. Conclusions

Monitoring the distribution of Cs in a H$^-$/D$^-$ ion source is of paramount importance to maximize intensity, uniformity and duration of the extracted negative ion beam. The design of a LAS diagnostic for the SPIDER experiment has been presented: 4 lines of sight will probe Cs density in parallel to the PG, at different vertical heights and at different distances from it. A similar diagnostic has been built and successfully operated on the test bed for the SPIDER Cs ovens: the diagnostic allowed to get measurements of Cs density in the range 10$^{13}$ m$^{-3}$-10$^{16}$ m$^{-3}$. Minor issues related to absorption line saturation have been studied and solved. The effects of ground state depopulation have been studied, not just in terms of lowering of the measured Cs density but also in terms of modifications of the two groups of hyperfine transitions. A correction of the measurements based on the inhomogeneous Doppler broadening model improved the results but not allowed a complete removal of the systematic deviation. More tests at different values of laser beam intensity and Cs density will be performed in future. In any case, during the operation of SPIDER the laser beam intensity will be kept as low as possible, compatibly with the issue of line saturation.

## Acknowledgments

The work leading to this publication has been partially funded from Fusion for Energy under the contract F4E-OFC-531-1. This publication reflects the views only of the authors, and F4E cannot be held responsible for any use which may be made of the information contained therein. The views and opinions expressed herein do not necessarily reflect those of the ITER organization.

## References

[1] R. Hemsworth et al., Status of the ITER heating neutral beam system, Nuclear Fusion 49 (2009) 045006.

[2] U. Fantz et al., Physical performance analysis and progress of the development of the negative ion RF source for the ITER NBI system, Nucl. Fusion 49 (2009) 125007.

[3] U. Fantz et al., Development of negative hydrogen ion sources for fusion: Experiments and modelling, Chemical Physics 398 (2012) 7–16.

[4] P. Franzen et al., Progress of the ELISE test facility: results of caesium operation with low RF power, Nuclear Fusion 55 (2015) 053005.

[5] D. Marcuzzi, Detail design of the beam source for the SPIDER experiment, Fusion Engineering and Design 85 (2010) 1792–1797.

[6] R. Pasqualotto et al., A suite of diagnostics to validate and optimize the prototype ITER neutral beam injector, JINST 12 (2017) C10009.

[7] U. Fantz et al., Optimizing the laser absorption technique for quantification of caesium densities in negative hydrogen ion sources, Journal of Physics D: Applied Physics. 44 (2011) 335202.

[8] A. Rizzolo, Caesium oven design and R&D for the SPIDER beam source, Fusion Engineering and Design 88 (2013) 1007– 1010.

[9] R. Pasqualotto et al., Plasma light detection in the SPIDER beam source, contributed paper of SOFT '18 conference.

[10] S. Briefi et al., Correction factors for saturation effects in white light and laser absorption spectroscopy for application to low pressure plasmas, Physics of Plasmas 19 (2012) 053501.

[11] F.T. Arecchi, F. Strumia and H. Walther, Advances in Laser Spectroscopy, NATO Advanced Science Institutes (ASI) Series B. Vol. 95 (1983).

[12] D. A. Steck, Cesium D line data, available online at https://steck.us/alkalidata/cesiumnumbers.pdf (consulted

---

*author's email: marco.barbisan@igi.cnr.it*

on April 2019).

[13] A. Mimo, Diagnostics for negative ion source NIO1, Master′s thesis, Università degli studi di Padova (2014).

___________________________________________

*author′s email: marco.barbisan@igi.cnr.it*